\documentclass[fleqn,usenatbib]{mnras}

\usepackage[T1]{fontenc}

\DeclareRobustCommand{\VAN}[3]{#2}
\let\VANthebibliography\thebibliography
\def\thebibliography{\DeclareRobustCommand{\VAN}[3]{##3}\VANthebibliography}

\usepackage{graphicx}	
\usepackage{amsmath}	
\usepackage{amssymb}	

\usepackage{newtxtext,newtxmath}

\title[Flaring Latitudes]{Flaring Latitudes in Ensembles of Low Mass Stars}

\author[E. Ilin et al.]{
Ekaterina Ilin,$^{1}$\thanks{E-mail: eilin@aip.de}
Ruth Angus,$^{2,3}$
Rodrigo Luger,$^{3}$
Brett M. Morris,$^{4}$
Florian U. Jehn,$^{5,6}$
\\
$^{1}$Leibniz Institute for Astrophysics Potsdam (AIP), An der Sternwarte 16, 14482 Potsdam, Germany\\
$^{2}$Department of Astrophysics, American Museum of Natural History, 200 Central Park West
New York, NY 10024, USA\\
$^{3}$Center for Computational Astrophysics, 162 5th Ave., New York, NY 10010, USA\\
$^{4}$Space Telescope Science Institute, Baltimore, MD 21218, USA\\
$^{5}$Institute for Landscape Ecology and Resources Management (ILR),\\Research Centre for BioSystems, Land Use and Nutrition (iFZ),\\Justus Liebig University Giessen, Heinrich-Buff-Ring 26, 35390 Giessen, Germany\\
$^{6}$Alliance to Feed the Earth in Disasters (ALLFED), Fairbanks, AK, USA.
}

\date{Accepted 2023 May 31. Received 2023 May 30; in original form 2022 December 21}

\pubyear{2023}

\begin{document}
\label{firstpage}
\pagerange{\pageref{firstpage}--\pageref{lastpage}}
\maketitle

\begin{abstract}
The distribution of small-scale magnetic fields in stellar photospheres is an important ingredient in our understanding of the magnetism of low mass stars. Their spatial distribution connects the field generated in the stellar interior with the outer corona and the large scale field, and thereby affects the space weather of planets. Unfortunately, we lack techniques that can locate them on most low-mass stars. One strategy is to localize field concentrations using the flares that occur in their vicinity.

We explore a new method that adapts the spot simulation software \texttt{fleck} to study the modulation of flaring times as a function of active latitude. We use empirical relations to construct flare light curves similar to those available from Kepler and the Transiting Exoplanet Survey Satellite (TESS), search them for flares, and use the waiting times between flares to determine the location of active latitudes. 

We find that the mean and standard deviation of the waiting time distribution provide a unique diagnostic of flaring latitudes as a function of the number of active regions. Latitudes are best recovered when stars have three or less active regions that flare repeatedly, and active latitude widths below 20 deg; when either increases, the information about the active latitude location is gradually lost. We demonstrate our technique on a sample of flaring G dwarfs observed with the Kepler satellite, and furthermore suggest that combining ensemble methods for spots and flares could overcome the limitations of each individual technique for the localization of surface magnetic fields.
\end{abstract}

\begin{keywords}
stars: magnetic field -- stars: flare -- methods: statistical
\end{keywords}



\section{Introduction}
The presence and location of active latitudes -- belts of strong, small-scale surface magnetic fields, and elevated stellar activity -- is poorly known on stars other than the Sun~\citep{berdyugina2005starspots, hathaway2015solar}. Disk-integrated observations that measure stellar activity of the star as a whole, e.g., chromospheric and coronal emission, spot-induced variability, and flares, have provided crucial constraints on how stars produce magnetic fields~\citep{kovari2014observing}, and their impact on habitability~\citep{airapetian2020impact}. However, to further advance these fields, spatially resolved information about various stellar activity phenomena -- like active latitudes of starspots and flares -- is required, yet largely missing. 

\subsection{Stellar dynamo}
The amplification of magnetic fields inside the Sun, a process known as solar dynamo, is driven by the movement of hot plasma in the convection zone, and forces introduced by stellar differential rotation~\citep{parker1955hydromagnetic}. Once the field is amplified, it emerges through the surface, forming complex magnetic loop structures. Their footpoints form active regions and sunspots in a range of low latitudes, which vary throughout the solar cycle to form the famous butterfly diagram~\citep{hathaway2015solar}. The stellar counterparts of these spots and flares can be large enough to be detected in disk-integrated observations. We observe them as modulations of the total stellar brightness periodic with stellar rotation, or sudden brightenings during flares~\citep[e.g.][]{notsu2013superflares}. 

In low mass stars, the exact mechanism behind the stellar dynamo is under active debate~\citep{brun2017magnetism}. The interior of a star can be measured using stellar oscillations, but asteroseismology is currently limited to bright solar-type and giant stars~\citep{garcia2019asteroseismology, rodriguez-lopez2019quest}, so that observers have to mostly rely on indicators from the exterior. One way to discriminate between different models is to map where magnetic fields emerge from the stellar interior to the surface~\citep{fan2009magnetic, bice2022longitudinally, weber2016modeling}. The locations of these dynamic fields are pinpointed by stellar flares, magnetically driven explosions in the corona that heat the footpoints to emit thermally at $\sim 10,000\,$K~\citep{kowalski2013timeresolved}, which can found in all stars with an outer convection zone~\citep{davenport2016kepler, gunther2020stellar, howard2019evryflare}. Flaring latitudes are of particular interest because they are direct evidence of dynamic and strong small-scale surface fields, a marker of field emergence~\citep{benz2010physical}. For example, while old Sun-like stars are expected to show low latitude active regions~\citep{fan2009magnetic}, models suggest that in young rapidly rotating stars they should arise closer to the poles~\citep{schuessler1992why,yadav2015formation, weber2016modeling}. 
Moreover, active latitudes shift over the course of the solar cycle, another distinct feature of the dynamo operating within, which is suggested in other stars as well~\citep{nielsen2019starspot}. 

\subsection{Space weather}
When flares are accompanied by particle eruptions with a preferred direction, such as coronal mass ejections (CME) and energetic particle events, their launch site influences whether a planet in the orbit will be hit or missed by the blast. If the planet is hit by a fast cloud of energetic and magnetized plasma, the planet can experience both temporary and permanent changes in its atmospheric chemistry~\citep{tilley2019modeling, chen2021persistence,chadney2017effect}, as well as increased atmospheric escape~\citep{hazra2022impact}. If the magnetic activity is as high as in many of the cool stars observed to date, the planet may even suffer a partial or complete loss of its envelope over time~\citep{khodachenko2007coronal,lammer2007coronal,cherenkov2017influence}. Therefore, tracing flaring latitudes could explain variability in exoplanet atmosphere characteristics, such as, for instance, conjured for HD 189733 b. Its atmospheric hydrogen appears to be escaping at lower rates before a flare was observed in X-ray compared to afterward~\citep{lecavelierdesetangs2012temporal}. If this is the case, the associated particles likely erupted from a latitude near the planet's orbital plane.

\subsection{Exoplanet characterization}
Even if the effects of latitude-dependent stellar activity are moderate for the planet, the presence of active latitudes indirectly complicates its characterization. The emission from a stellar surface with active latitudes is inhomogeneous. If it is then falsely assumed to be uniform, the depth of a transiting planet may be measured against a biased baseline that is brighter or darker than the average surface, and falsely yield a too large or too small radius for the planet~\citep{morris2018robust}. Furthermore, the presence of active regions in or out of the passageway of transiting planets can distort atmospheric signal in transmission spectra~\citep{mccullough2014water,rackham2018transit,cauley2017decade,cauley2018effects}, and interfere with the orbital radial velocity signal~\citep{huber2009longterm,meunier2019activity,meunier2019activitya}. For example, the characterization of the seven planets surrounding TRAPPIST-1 may be affected by an active latitude of bright spots associated with flaring activity outside their transit chords~\citep{morris2018possible,ducrot201884}. Without a solid understanding of the spatial distribution of active regions on the surfaces of exoplanet hosts, access to the planets' atmospheres will remain limited.

\subsection{Measuring active latitudes}
On the Sun, flares and spots occur in a belt below about 30 deg latitude around the equator~\citep{chen2011statistical}. On other stars, we know comparably little. Active latitudes have been measured in a number of ways, including spot occultations by misaligned planets~\citep{morris2017starspots,netto2020stellar}, asteroseismology~\citep{thomas2019asteroseismic, bazot2018butterfly}, light curve inversions of stars with rotational modulation by spots~\citep{huber2010planetary, santos2017starspot, basri2022new}, and, albeit with low spatial resolution, using spectropolarimetry~\citep{unruh1995evolution,perugini2021evolution}. Some of the handful of stars investigated in these works appear active, with spots located close to the poles or preferably near the equator. Others show spots at multiple distinct latitudes, or in a broad range, but the sample is too small to confidently identify meaningful trends. 

Localizing flaring latitudes is particularly challenging. They occur randomly in time, limiting time-resolved spectroscopic methods to the most actively flaring nearby stars, such as BO Mic~\citep{wolter2008doppler}. Except for the Sun, we lack the observations to tell if and how they are spatially related to spots, which would otherwise simplify the task. Stellar flare locations have only been captured serendipitously~\citep[e.g.][]{wolter2008doppler, schmitt1999continuous}, or for exceptionally energetic events on particularly rapidly rotating stars~\citep{ilin2021giant}. For the bulk of flaring stars, however, localization methods are missing. 

This work explores a method that allows us to systematically infer the flaring latitudes in ensembles of a wide range of stars using photometric observations, such as currently available from the Kepler~\citep{koch2010kepler} and Transiting Exoplanet Survey Satellite~\citep[TESS,][]{ricker2015transiting} missions. In Section~\ref{sec:ensembles:methods}, we present our simulations of flaring light curves on stars with active latitudes. In Section~\ref{sec:ensembles:results}, we use these synthetic light curves to show how we can use the occurrence times of flares to recover the location of an active latitude, and apply the method to a sample of flaring G dwarf stars in Kepler. We discuss the advantages and limitations of this approach in Section~\ref{sec:ensembles:discussion}, and summarize in Section~\ref{sec:summary}. 

\section{Methods}
\label{sec:ensembles:methods}
We simulate ensembles of flaring stars to explore if information about the latitudinal distribution of flaring regions can be inferred from the waiting time distributions of flares in stellar ensembles. 

\subsection{Waiting time and night length}

We define \textbf{waiting time} as the time between two subsequent flares phase folded with stellar rotation. That is, waiting time is measured in units of rotational phase, not days, allowing us to build ensembles from stars with varying rotation periods. In other words, we measure the flare phases, sort them ascending, and calculate the waiting times from the resulting series of timestamps.

\begin{figure}
    \centering
    \includegraphics[width=0.9\hsize]{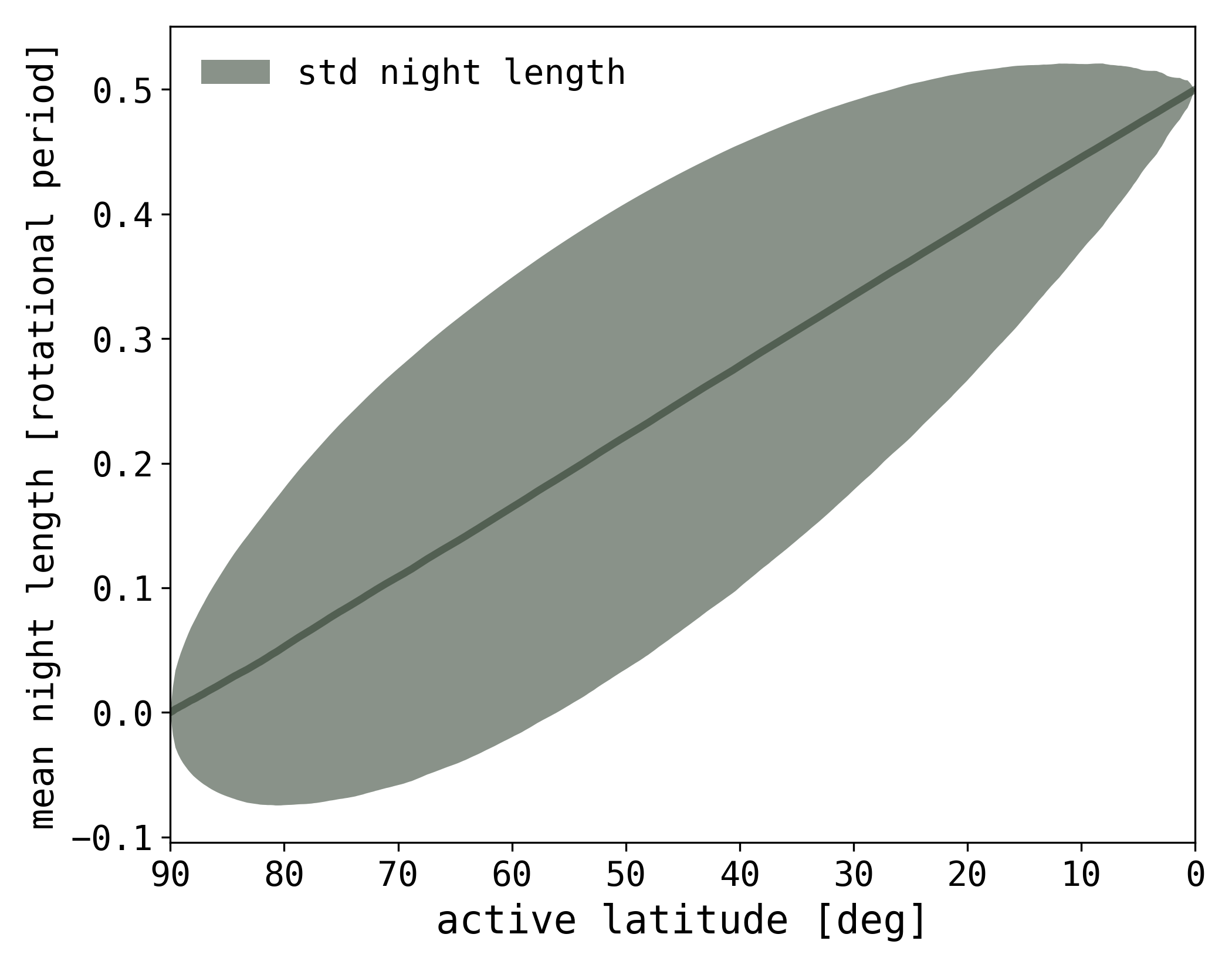}
    \caption{Mean and standard deviation of night lengths in an ensemble of randomly oriented stars. We define night length as the time during which a flaring region is hidden from the observer as the star rotates. If the active latitude is positioned either at the pole or the equator, the orientation does not affect the night length. Otherwise, night lengths vary with inclination~(grey shaded area).}
    \label{fig:nightlength}
\end{figure}

A flaring region viewed equator-on, i.e. with stellar inclination $i=90^\circ$, will be seen flaring $50\%$ of the time, and will be hidden behind the limb for the other $50\%$. In contrast, a flaring region close to the rotational pole with $i=0^\circ$ is seen $100\%$ of the time. At all other inclinations, the night length will vary with the latitude of the active region~(Fig.~\ref{fig:nightlength}). Here, \textbf{night length} is defined as the time during which a flaring region is hidden from the observer, given its latitude and the star's inclination. For an ensemble of stars with random orientations, we observe a distribution of night lengths with a characteristic mean and standard deviation. We can trace the night length through the timing of flares that occur in the active latitude, which allows us to infer its location. As our results will show, the picture is somewhat more complicated, which we address in Section~\ref{sec:ensembles:results:realistic}.

\subsection{Marginalizing over inclination}

For the stars in the ensemble we do not know the inclination $i$ which, for any individual star, creates a degeneracy in the latitude of a particular flaring region. However, we can assume that the stars' rotation axes are oriented randomly, so that we can marginalize over $i$. If the flaring regions in the ensemble are located at approximately the same latitude, the night lengths and therefore the waiting times of the ensemble will have a distribution characteristic of that latitude. 

\subsection{Ensembles of flaring stars}
We simulate light curves of ensembles of flaring stars with different active latitudes, and apply standard flare finding techniques to identify flare peak times, i.e., iteratively clipping positive outliers from a rolling median in the light curve to find the quiescent flux level, and then identifying series of outliers from this quiescent level as flare candidates. We then calculate the waiting time distribution, and use the mean $\mu$ and standard deviation $\sigma$ of the distribution to characterize it as a function of active latitude. In the simple case of a well-localized flaring region at a fixed active latitude $\theta$, we construct the ensemble as follows:

For each star, we define a flaring region, assuming that it is stable. Stable means that in the course of observation, the region has already fully emerged, does not decay, and produces flares randomly, following a Poisson process in time at a defined rate. In the application to real observations, the varying lifetimes of spots and active regions need to be taken into account~(see Sections~\ref{sec:results:okamoto}~and~\ref{sec:ensembles:discussion:lifespots}). 

\subsection{Individual flares}
\label{sec:methods:indflare}

\begin{figure}
    \centering
    \includegraphics[width=\hsize]{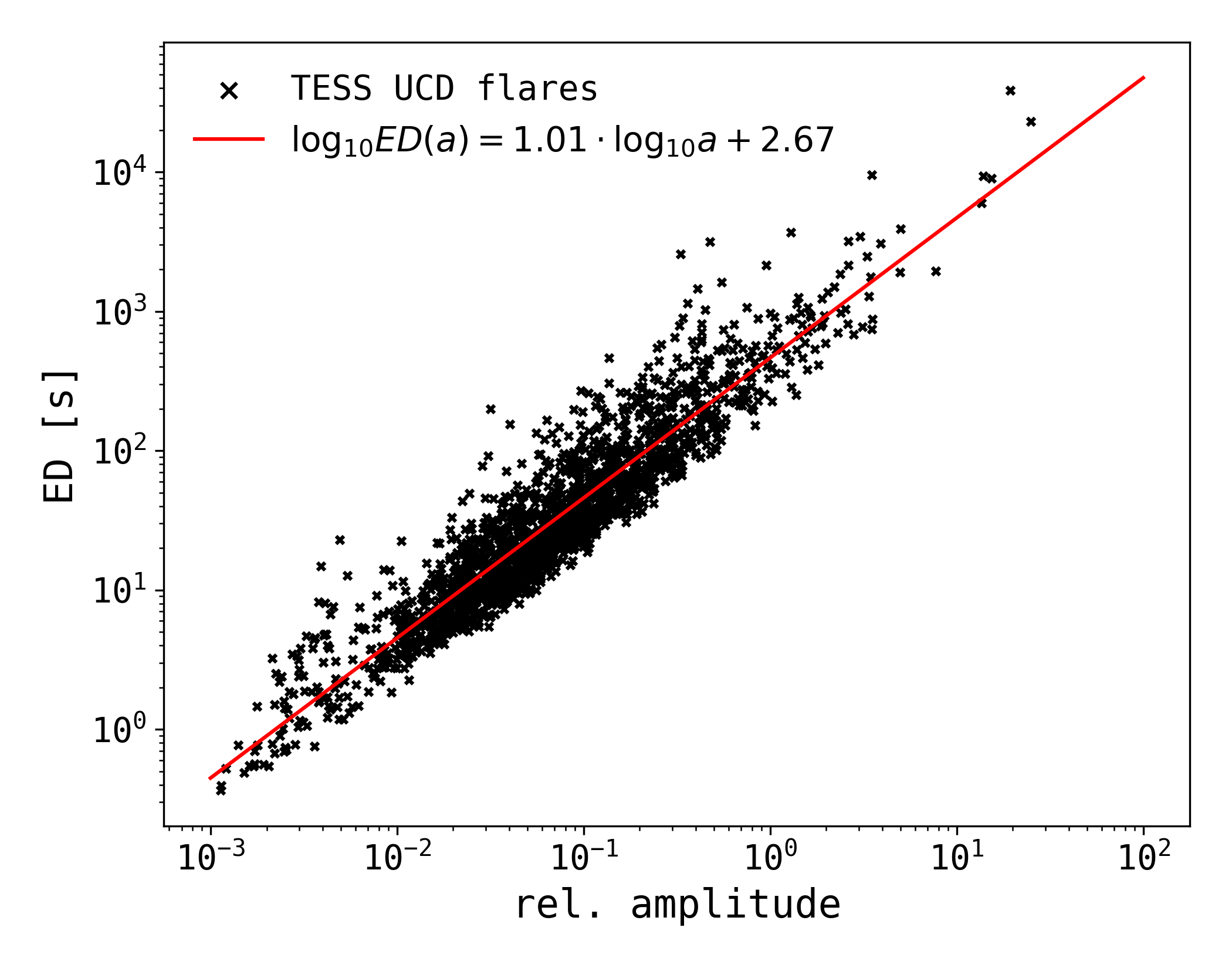}
    \caption{The relation between equivalent duration $ED$ and relative amplitude of flares (black dots) in ultra-cool dwarfs (spectral type M6-M9) observed by TESS (Pineda et al. in prep.). The red line is the power law fit used to infer relative amplitude $a$ from $ED$~(see~Section~\ref{sec:methods:indflare}).}
    \label{fig:ucdamplitudes}
\end{figure}

To create a realistic individual flare, we invert the~\citet{davenport2014kepler} flare template parametrization with variable full width at half the maximum flux ($t_{1/2}$) and amplitude ($a$). We infer $a$ from the empirical relation between amplitude and equivalent duration $ED$ using a sample of ultracool dwarfs (Fig.~\ref{fig:ucdamplitudes}). The equivalent duration $ED$ is the time the star needs to emit as much light as the flare did~\citep{gershberg1972results}, which is a relative unit for the flare energy. With given $ED$, we infer $a$, and then use the parametrization from \citet{davenport2014kepler} to calculate $t_{1/2}$. 

\subsection{Flare frequency distributions}

Since flares are distributed on a power law across a wide range of energies $E$ (or $ED$) and stellar properties~\citep{shibayama2013superflares, maehara2015statistical, ilin2021flares}, we create a number $N$ of flares based on flare rate $\beta$, drawn from an energy distribution with power law slope $\alpha$:

\begin{equation}
    N(E) = \beta \cdot E^{-\alpha}
    \label{eq:powerlaw}
\end{equation}

In our simulations, the distribution of flare energies in each flaring region follows a power law with slope $\alpha$ between 1.5 and 2.5, and a rate $\beta$ between 1 and 60.

\subsection{Flare injection, rotational modulation, and recovery}
\label{sec:methods:injrec}
\begin{figure}
    \centering
    \includegraphics[width=0.9\hsize]{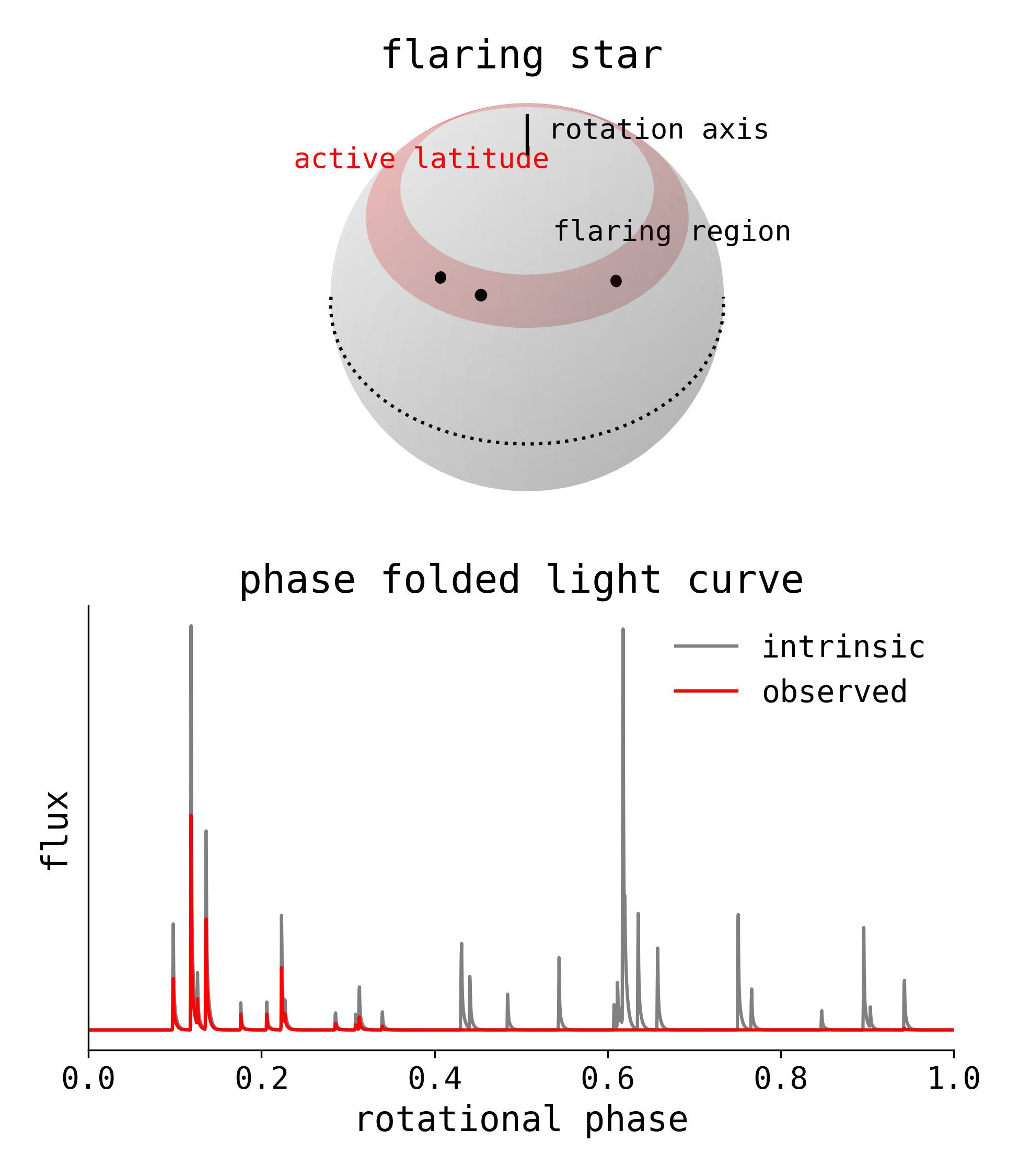}
    \caption{Model illustration: Star with flaring active latitudes. Top: stellar sphere with three active regions placed within an active latitude strip. Bottom, grey line: Phase folded light curve of a single active region, assuming other variability signatures have been completely removed. Bottom, red: Same light curve but as seen by the observer, accounting for geometric foreshortening and visibility, as the star rotates. As a consequence, some flares appear lower in amplitude, while others are not detected at all~(see Section~\ref{sec:methods:injrec}).}
    \label{fig:modelillustration}
\end{figure}

We then distribute the flares with their respective $a$ and $t_{1/2}$ in a light curve according to a Poisson process (grey curve in Fig.~\ref{fig:modelillustration}). We chose the light curve cadence and Gaussian noise such that all injected flares are recovered in the unmodulated light curve.  That means that all flares have their peaks $3\sigma$ above the noise level for at least three consecutive data points. Next, we assign a latitude $\theta$ and arbitrary longitude $\phi$, pick a random orientation of the star (random in $\cos i$). We use \texttt{fleck}, a software package for simulating rotational modulation of stars due to starspots~\citep{morris2020fleck,morris2020relationship} to rotate the flaring stars, replacing the dark spots with flaring regions. In other words, in \texttt{fleck}, we replace the constant darkening caused by a spot with the time-varying brightening caused by a flare. We fix the flaring region area at $\sim10^{-4}$ of the stellar hemisphere area, or equivalently $1\%$ of the stellar radius, which is within the $10^{-5}-10^{-3}$ range of estimated areas of the white light footpoints of energetic solar and stellar flares~\citep{metcalf2003trace, kowalski2010white, kowalski2013timeresolved}. For completeness, we applied quadratic limb darkening. This only marginally affects the flare amplitude compared to geometric foreshortening, assuming that the flare emission is optically thick~(see Fig.~B1 in \citealt{ilin2021giant}). 

We do not apply spot induced rotational modulation to the light curve, assuming that it was removed successfully while preserving the flare signal, for which various algorithms exist~\citep[e.g.][]{shibayama2013superflares, vandoorsselaere2017stellar, raetz2020rotationactivity, ilin2021flares}. These algorithms can distinguish rotational modulation from flares as long as the time scales are sufficiently different for both. This is true even for the most rapidly rotating stars with periods below $10\,$h, where very long duration flares can occur, but are very rare~\citep{ilin2021giant}. Those events are easily identified as flares by eye due to their large amplitudes. Flares that are shorter, but still evolve more gradually than typical fast-rise-exponential-decay flares~(see~Fig.~\ref{fig:modelillustration}) may be lost while removing rotational modulation. But for those, the distinction between flares and other sources of variability like spots or faculae becomes increasingly uncertain, so it is not clear if missing these "flares" is an actual loss. 

After rotating the star, a fraction of flares is no longer detected with \texttt{AltaiPony} because it occurs on the back of the star, or close to the limb, so that the amplitude falls below the noise level (red curve in Fig.~\ref{fig:modelillustration}). We only use these detected flares to calculate the waiting times.

\subsection{Ensembles of stars with different active latitudes}

We repeat the above steps 200 times for each choice of latitude $\theta$. This creates an ensemble of 200 flaring light curves on stars with random orientations, random flare times, picking $\alpha$ randomly from 1.5 to 2.5, choosing random flaring rates $\beta$. All stars share the same flaring latitude with the same width ($\Delta\theta=5\,$deg). From simulation to simulation, we vary the average $\beta$ in the ensembles so that our results cover different mean waiting times $\mu$~(see~Fig.~\ref{fig:results}). We note that we do not need to specify the number of rotation periods covered by each light curve, because we only require phase-folded light curves to measure the waiting times. At a fixed flare rate, more rotations mean more flares in the total light curve, which we can control by setting $\beta$ directly. 

The number of 200 stars in the sample is motivated, on the one hand, by the requirement that we need a sufficient number of stars to marginalize over inclination. On the other hand, we wanted to demonstrate our technique on a realistic number of flaring stars with similar properties, and therefore, expectedly similar flaring latitudes~(see Section~\ref{sec:ensembles:discussion} for a discussion of relevant criteria in the sample selection). This can be achieved, for instance, with the growing archive of space based light curves of flaring stars~\citep[see, e.g.,][]{davenport2016kepler,gunther2020stellar, yang2017flaring, feinstein2020flare}.  

We list the studied ranges for $\alpha$, $\beta$, and further parameters in Table~\ref{tab:modelparams}.

\begin{table}

    \caption{Model parameters.}
    \begin{tabular}{l|r}
         parameter & values \\\hline\hline
         flare rate $\beta$&1-60 flares per light curve\\
         energy distribution power law slope $\alpha^*$&$1.5-2.5$\\
         number of flaring regions per star & 1-5 \\
         active latitude center $\theta$ & 5-85 deg\\
         active latitude width $\Delta \theta$ & 5-40 deg\\
         active longitude $\phi$ & 0-360 deg random\\
         flaring region size & $10^{-4}$ stellar hemisphere\\
         quadratic limb darkening coefficients & 0.5079, 0.2239\\
         stars in ensemble & 200 \\
    \end{tabular}
    \newline
    $^*$ per active region
    \label{tab:modelparams}
\end{table}

\section{Results}
\label{sec:ensembles:results}
Based on the results of our simulations, we calibrate quadratic relations between $\theta$ and the mean and standard deviation of waiting times, for different numbers of flares and different hemispheric configurations. The only additional information needed about the stars is their rotational period. This process is outlined in more detail in Section~\ref{sec:ensembles:results:analytic}.

We can use the waiting times between flares to locate the active latitude on the stellar surface if an actively flaring latitude is present in an ensemble of stars. Our results present a straightforward way to check whether the waiting time distribution of flares is informative in a given ensemble, and to find out what it can tell us about the presence, number and location of active latitudes. Our simulations were chosen to represent the currently available sample sizes achievable with light curves from the Kepler and TESS archives. While compiling a sufficiently large data set is outside the scope of this paper, in Section~\ref{sec:results:okamoto}, we demonstrate how our technique can be applied using a small sample of flaring G dwarfs observed by Kepler~\citep{okamoto2021statistical}.

\subsection{Analytical expression for active latitude inference}
\label{sec:ensembles:results:analytic}
\begin{figure}
    \centering
    \includegraphics[width=\hsize]{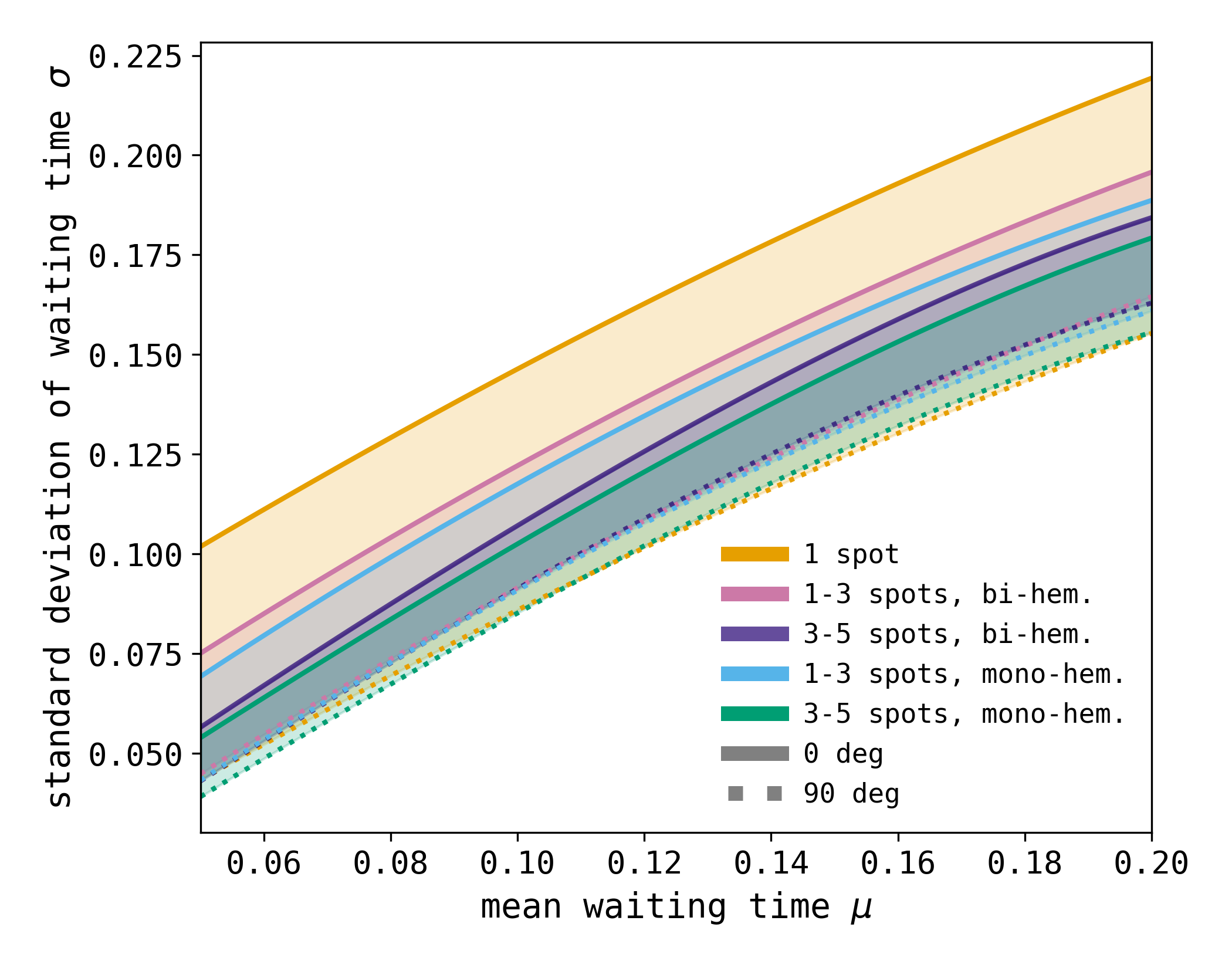}
    \caption{Ranges of active latitudes as a function of $\mu$ and $\sigma$, calculated from the relations in Table~\ref{tab:fit}. Each line shows the $\theta=0$ deg (solid) and $90$ deg (dotted) margins of the colored latitude range in between. Stellar ensembles with a single low latitude active region occupy a distinct region in the diagram (orange area). Other combinations of $\mu$ and $\sigma$ are ambiguous because two or more ranges overlap.} 
    \label{fig:results}
\end{figure}

\begin{figure}
    \centering
    \includegraphics[width=\hsize]{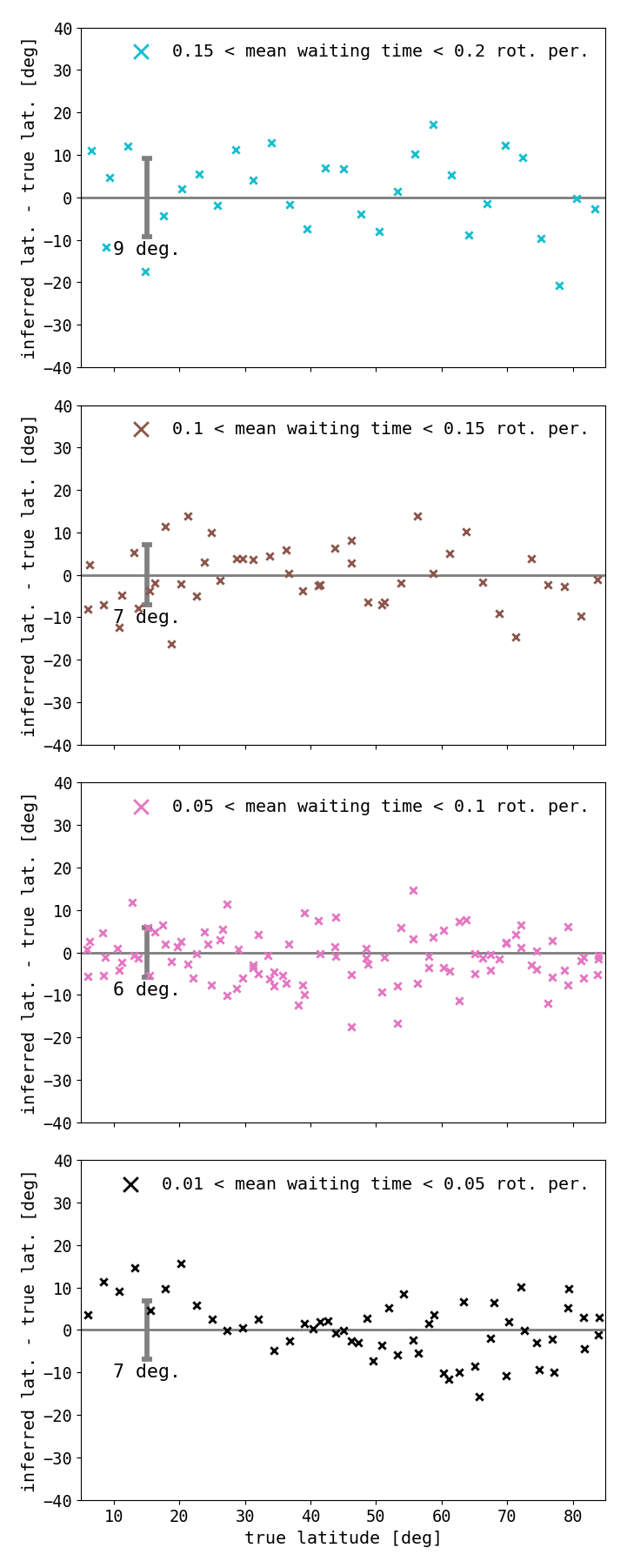}
    \caption{Residuals of inferred latitudes as a function of true latitude. Each data point represents a set of 200 stars with one flaring region placed randomly on either hemisphere, but in the same latitude range for all stars. The residuals decrease with increasing flare rate, i.e., lower $\mu$. The top panel shows the subset with the lowest flaring rate, the bottom panel shows that with the highest flaring rate. To the left in each panel, we show the standard deviation in the residuals.} 
    \label{fig:results:1spot}
\end{figure}

\begin{figure}
    \centering
    \includegraphics[width=\hsize]{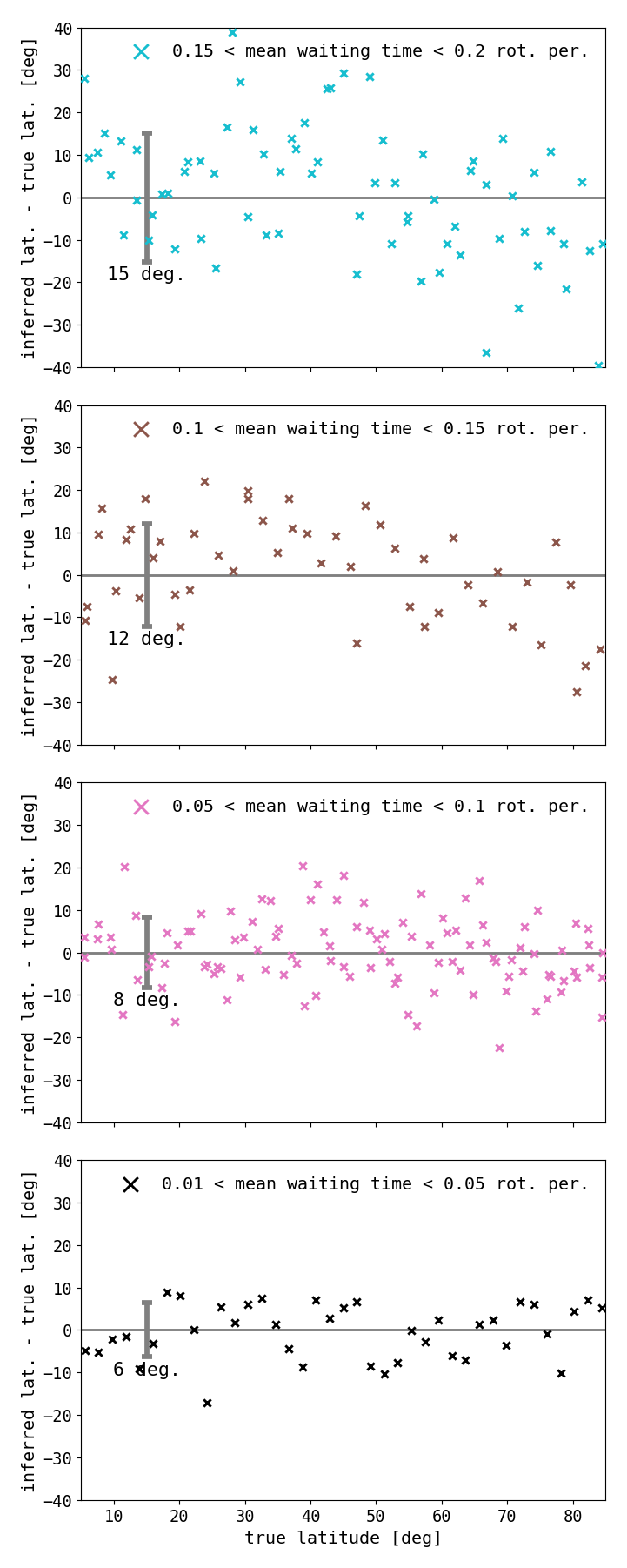}
    \caption{Same as Fig.~\ref{fig:results:1spot}, but for 200-star ensembles with 1-3 flaring regions distributed on both hemispheres.} 
    \label{fig:results:13spots}
\end{figure}

In our chosen set of simulations, we picked a $\Delta \theta =5$ deg wide active latitude, a typical solar value around the activity maximum~(Fig.~32 in \citealt{hathaway2015solar}), and explored the cases of either an ensemble of stars with only one active region per star, or 1-3, or 3-5 actively flaring regions per star. In the case of 1-3 and 3-5 flaring regions, we distinguish the cases where either all flaring regions are on the same hemisphere (mono-hemispheric) or distributed over both hemispheres (bi-hemispheric), at random longitudes, because the night lengths for stars with inclined rotation axes differ for Northern and Southern hemispheres. On the Sun, active regions appear on both hemispheres. However, this is poorly constrained on other stars. In magnetohydrodynamical simulations of fully convective M dwarfs, flux emergence can potentially be both mono- and bi-hemispheric~\citep{browning2008simulations, brown2020singlehemisphere, kapyla2021starinabox}. We therefore simulate light curves for eight runs of 100,000 light curves each for each of the five setups (1, 1-3 mono-hem., 1-3 bi-hem., 3-5 mono-hem., 3-5 bi-hem.). The eight runs differ by the number of flares chosen for each active region, which translates into a range of detected flaring rates. We then split the sample into ensembles of 200 stars with the same active latitude $\theta$ and calculate $\mu$ and $\sigma$ for the ensemble, taking 5 deg steps in latitude, and ignoring the equator and the pole, i.e., $[5,10,..,85]$ deg. We fit a polynomial expression, 

\begin{equation}
    \theta = a_1\cdot\mu^2 + a_2\cdot\mu + b_1\cdot\sigma^2 + b_2\cdot\sigma + c,
    \label{eq:fit}
\end{equation}
to the resulting (eight runs) $\times$ (17 latitudes) $=$ 136 data points in each setup. The best-fit parameters are given in Table~\ref{tab:fit}.

The range of resulting $\mu$ and $\sigma$ values in our synthetic data is shown in Fig.~\ref{fig:results}. The colored areas contain the measured combinations of $\mu$ and $\sigma$ for each setup. That means, if a combination of $\mu$ and $\sigma$ measured in real observations falls outside the colored areas, one or more of the assumptions of the simulation must be violated, e.g., the assumption that flares are generated randomly in time in any given flaring region~(see Section~\ref{sec:methods:injrec}, and the discussion in Section~\ref{sec:ensembles:discussion:poission}). 

We validated the fitting results using additional simulations of stellar ensembles that were not used for the fit. Figs.~\ref{fig:results:1spot} and \ref{fig:results:13spots} show the residuals when using the best-fit parameters in Table~\ref{tab:fit} to infer $\theta$ in the 1-spot and the bi-hemispheric 1-3 spot cases, respectively. The fewer active regions, the better the recovery of flaring latitudes, if the number of active regions is known. As expected, the uncertainty in the $\mu$-$\sigma$ relation for any given active latitude decreases with increasing average flare rate in the ensemble, or, equivalently, shorter mean waiting time $\mu$ (to the left in Fig.~\ref{fig:results}). A shorter mean waiting time can be achieved by increasing the observing baseline per star.

\subsection{Flaring region number and hemisphere distribution vs. latitude}
\label{sec:ensembles:results:ambiguities}
For most combinations of the mean $\mu$ and standard deviation $\sigma$ of the waiting time distribution, their interpretation is ambiguous in two ways. Except for single active regions at low latitudes (orange area in Fig.~\ref{fig:results}), multiple interpretations are usually possible. 

First, in the regions where the permitted ($\mu,\sigma$) areas overlap in Fig.~\ref{fig:results}, a solution with fewer spots at higher latitudes is degenerate with a solution with more spots at lower latitudes. For example, 1-3 active regions near the equator (solid magenta line in Fig.~\ref{fig:results}) overlap with the range for a single active region at mid-latitudes. 

\begin{figure}
    \centering
    \includegraphics[width=\hsize]{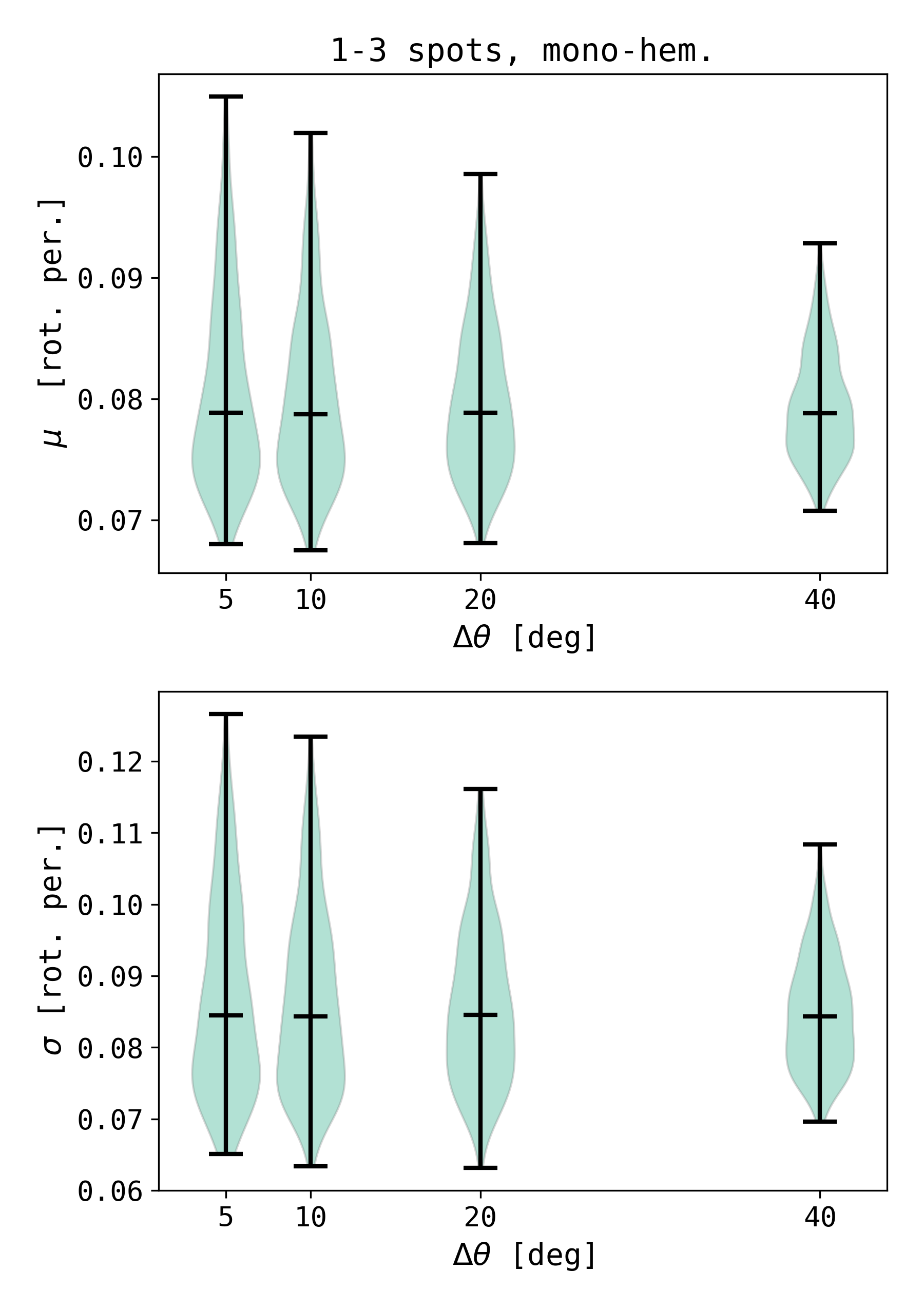}
    \caption{Distribution of mean $\mu$ (top panel) and standard deviation $\sigma$ (bottom panel) of waiting time distributions for four groups of simulated 200-star ensembles with varying width of the flaring latitude $\Delta\theta$ (i.e., width the of the red strip in Fig.~\ref{fig:modelillustration}). Each group is shown as a blue violin plot with minimum, maximum and median indicated as horizontal black lines. All other parameters are fixed: Each ensemble is composed of stars with 1-3 flaring spots on one hemisphere, and the same average intrinsic flaring rates. For narrow active latitudes, we see wider distributions in $\mu$ and $\sigma$ than for wider active latitudes. This is because narrower active latitudes close to the pole or to the equator produce more extreme waiting time distributions, described by $\mu$ and $\sigma$, than wider active latitudes.}
    \label{fig:vardeltatheta}
\end{figure}

Second, the wider the active latitude becomes, the more the information on latitude is lost. Fig.~\ref{fig:vardeltatheta} shows distributions of $\mu$ and $\sigma$ for four simulations side by side, where only the width of the active latitude $\Delta\theta$ varies between the simulations. For $\Delta\theta\approx10\,$deg and below, the results are completely unaffected, but wider $\Delta\theta\geq20\,$deg increasingly lose the latitudinal information. In the extreme case, where flaring regions are spread over the entire hemisphere ($\Delta\theta=90\,$deg), the waiting time distribution converges on the 45 deg line that roughly bisects each area in Fig.~\ref{fig:results}~(see also Fig.~\ref{fig:appendix}). We also investigated the distinction between mono- and bi-hemispheric distributions of active regions, and found that~ we cannot distinguish them in our data, so that our latitude inference does not depend on whether there is only one or two active latitudes on opposite hemispheres of the stars.

In practice, given some observed ($\mu,\sigma$) combination, one can use all available fits in Table~\ref{tab:fit}. Those that return a value between 0 and 90 deg are possible solutions for $\theta$, the remainder can be ruled out. In this way, the latitude inference constrains the latitude of the active region as a function of their number~(see Section~\ref{sec:results:okamoto} for a demonstration on real observations).

\subsection{Flare energies and amplitudes vs. latitude}
\begin{figure}
    \centering
    \includegraphics[width=\hsize]{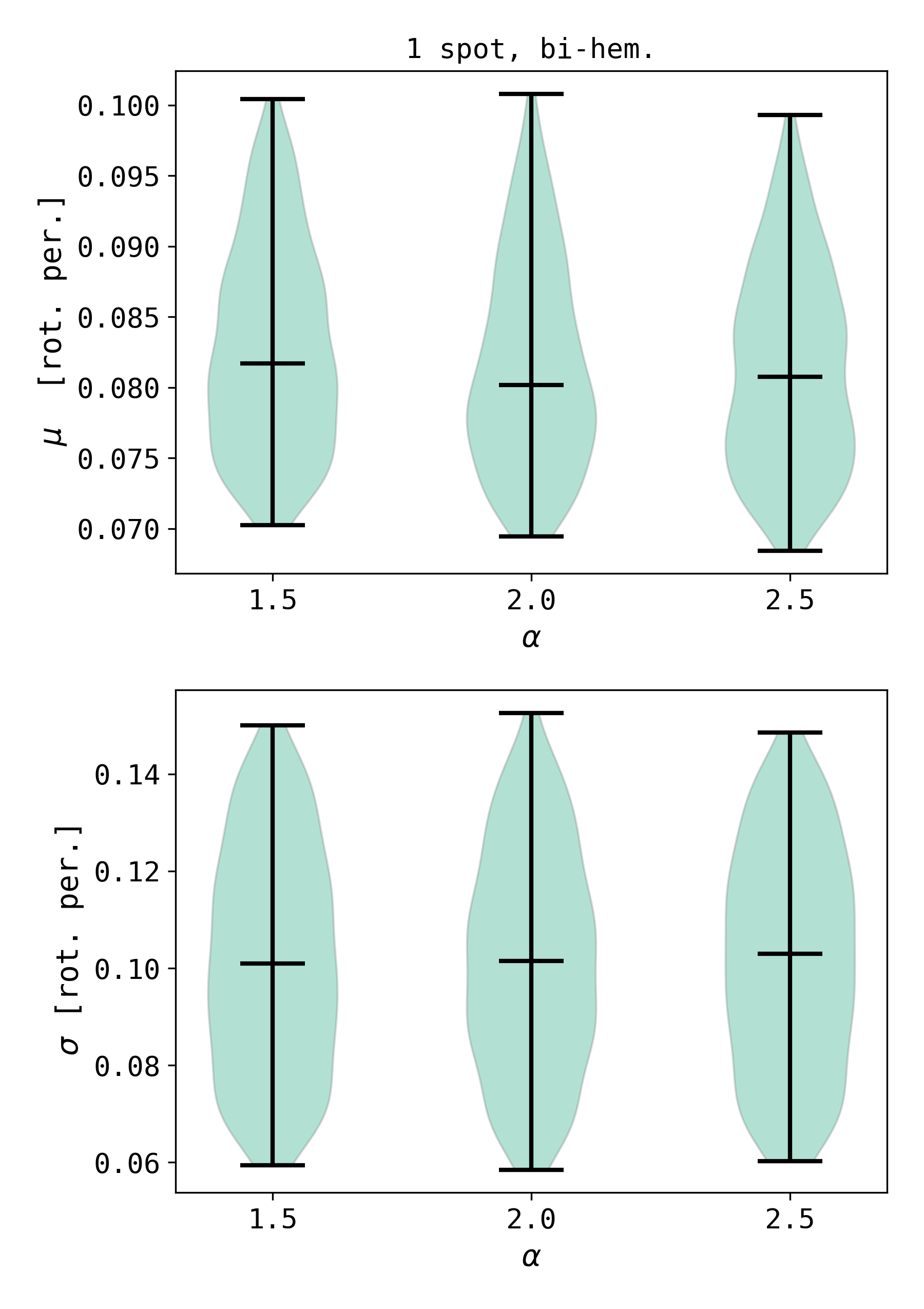}
    \caption{Same as in Fig.~\ref{fig:vardeltatheta}, now varying the power law slope~$\alpha$ of the flare frequency distribution in an empirically motivated range~\citep[see Fig.~13 in][for an overview]{ilin2021flares}. Each ensemble is composed of stars with a single flaring spot, and the same average intrinsic flaring rates. As opposed to Fig.~\ref{fig:vardeltatheta}, the waiting time distribution parameters $\mu$ and $\sigma$ (top and bottom panels) are not significantly affected by the changes in $\alpha$.}
    \label{fig:varalpha}
\end{figure}

The power law exponent of the flare frequency distribution and the placement of active regions on either one or both hemispheres (like on the Sun) have different effects on the latitude inference: First, a lower power law exponent of the flare frequency distribution by definition decreases the ratio of low to high amplitude flares~(see Eq.~\ref{eq:powerlaw}). This increases the number of detected flares close to the limb, where flares are geometrically most foreshortened, by increasing the average energy of flares. Both, the mean waiting time $\mu$ and the width of the waiting time distribution $\sigma$ should slightly decrease as an effect. However, Fig.~\ref{fig:varalpha} illustrates that this effect is insignificant in the investigated range of $\alpha$ values. Similarly, since amplitudes closely approximate flare energy~(Fig.~\ref{fig:ucdamplitudes}, or, similarly,~\citealt{hawley2014kepler}), varying flare amplitudes has a small effect on the waiting times, too. 

\begin{table*}
    \centering
    \caption{Best-fit parameters to Eq.~\ref{eq:fit} for all tested bi- and mono-hemispheric configurations with 1, 1-3, and 3-5 flaring regions (FR).}
    \begin{tabular}{lccccc}
\hline
       & 1 FR, bi-hem. & 1-3 FR, bi-hem. & 3-5 FR, bi-hem. & 1-3 FR, mono-hem. & 3-5 FR, mono-hem. \\
\hline
 $a_1$ &   $-1922 \pm 64$ &    $-4258 \pm 168$ &   $-16200 \pm 736$ &      $-6287 \pm 196$ &     $-13523 \pm 311$ \\
 $a_2$ &    $1606 \pm 20$ &      $3411 \pm 60$ &     $8536 \pm 269$ &        $4222 \pm 62$ &       $7356 \pm 111$ \\
 $b_1$ &     $577 \pm 95$ &      $390 \pm 325$ &     $9997 \pm 922$ &        $911 \pm 327$ &       $9425 \pm 456$ \\
 $b_2$ &   $-1623 \pm 25$ &     $-3023 \pm 80$ &    $-7674 \pm 273$ &       $-3564 \pm 73$ &      $-6957 \pm 121$ \\
   $c$ &       $84 \pm 1$ &         $65 \pm 2$ &         $16 \pm 3$ &           $47 \pm 1$ &           $14 \pm 1$ \\
\hline
\end{tabular}

    \label{tab:fit}
\end{table*}

\subsection{Realistic light curves are needed for the correct interpretation of flare waiting times}
\label{sec:ensembles:results:realistic}
To a first approximation, the trends observed in Fig.~\ref{fig:results} are the result of changing night length distributions~(Fig.~\ref{fig:nightlength}) for different active latitudes. At a fixed mean waiting time $\mu$ between flares, flaring regions placed at high latitudes would be visible for most orientations of the stellar rotation axis. Only a few regions would be hidden from the observer on average, so that the spread of waiting times $\sigma$ will be close to a 1-1 relation with $\mu$, as expected from the Poisson process assumption about when flares occur. So in principle, one could hope to find a closed-form solution. However, our method shows that there are observational effects that cannot be integrated in an analytical solution. We run a flare finding algorithm~\citep{ilin2021altaipony} on realistic light curves, where flares are geometrically foreshortened with increasing distance from the center of the stellar disk, are cut off at the ends of the light curve, and may randomly overlap. Most importantly, our maximum waiting time is capped at $1$ by definition because we measure waiting time in units of rotation period. This, in turn, decreases $\sigma$. The combined upshot of these effects is that, almost always, $\sigma\neq\mu$, even in the case of near pole-on active regions~(dotted lined in Fig.~\ref{fig:results}). 

\subsection{Application to flares on G-type Kepler stars}
\label{sec:results:okamoto}
It is out of the scope of this work to attempt a full-blown analysis of real observations, since that would require 1. careful and extensive sample selection, 2. flare finding and vetting, and 3. rotation period measurements on many stars, including corrections for completeness. However, for smaller samples of stars, analysis that includes several of these steps exists in the literature~\citep{howard2020evryflare,tu2021superflares, ramsay2020tess}. We have primarily focused on low-mass, and particularly M dwarfs, in this work~(see Fig~\ref{fig:ucdamplitudes}, and the flare model from~\citealt{davenport2014kepler}). However, the mid-M dwarf flare model is successfully used to describe flares on earlier spectral type stars~\citep{howard2022no, gunther2020stellar, jackman2018groundbased}, and the $ED$-amplitude relation is also comparable~\citep{liu2023minutecadence}. In this Section, we focus on G~dwarfs by using results of~\citet{okamoto2021statistical}, who searched all G~dwarf light curves in the Kepler archive for flares, and determined the stars' rotation periods. Since the Sun, a G dwarf, is the only star for which flaring latitudes are precisely known, a G dwarf ensemble is the only one with a reliable benchmark. 

We assume that given the multi-year baseline of Kepler observations for most of the stars in the sample, and the relatively short rotation periods ($P<40\,$d), the rotation phases of all stars are evenly covered by observations. We calculate the rotation phases from the flare times given in the online catalog\footnote{\url{https://cdsarc.cds.unistra.fr/viz-bin/cat/J/ApJ/906/72}, retrieved on March 27, 2023.} in \citet{okamoto2021statistical}. We only consider stars with 5-30 detected flares in the data, so that we stay within the limits of our simulations. This mostly removes stars with less than 5 flares, but also some active and bright stars with more than 30 flares, leaving 61 stars in the sample. Then we calculate the mean waiting time $\mu$ and standard deviation $\sigma$ in units of $P$ for the fast ($P<10\,$d) and slow ($P>10\,$d) rotators in the sample. We also further split the fast rotators into very fast ($P<5\,$d), and moderately fast rotators ($5<P<10\,$d). Within these four subsamples, we may expect that the number of flaring regions and their latitudes will be similar because their dynamos are strongly determined by stellar mass and rotation~\citep{brun2017magnetism}. In particular, around the $10\,$d mark, stellar activity changes from the saturated regime for fast rotators, where an increase in rotation speed no longer causes an increase in activity, to the unsaturated regime for slow rotators, where rotation and activity are correlated~\citep[e.g., in flaring activity,][]{davenport2016kepler, nunez2022factory, ilin2021flares}.

\begin{figure}
    \includegraphics[width=\hsize]{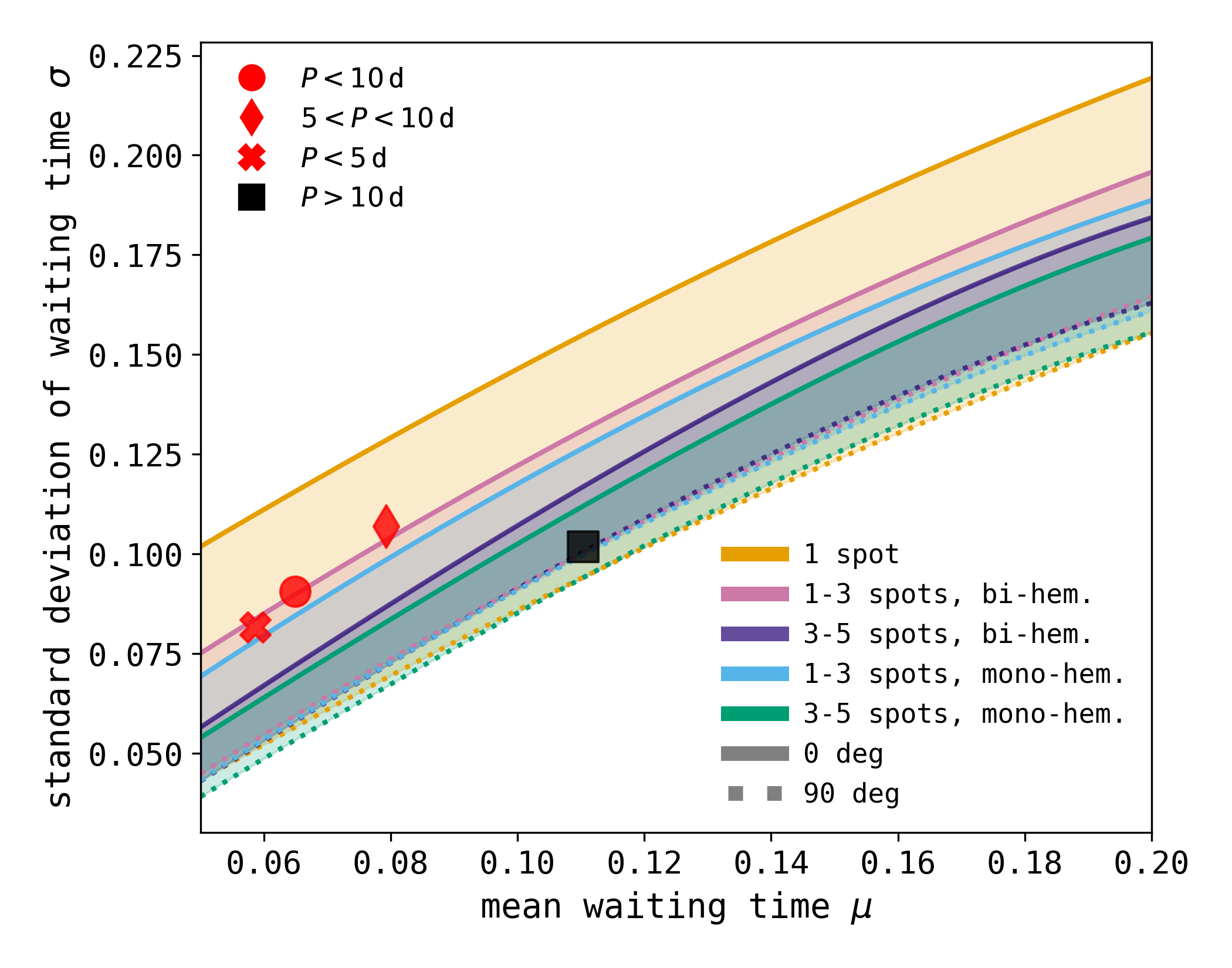}
\caption{Same as Fig.~\ref{fig:results}, but with added mean and standard deviation of flaring G dwarfs from~\citet{okamoto2021statistical} in different intervals of the rotation period $P$ (red and black symbols). See Table~\ref{tab:okamoto} for the sample parameters and results in each interval.}
    \label{fig:okamoto}
\end{figure}

In Table~\ref{tab:okamoto}, we show the numbers of stars and flares in each subsample along with the inferred latitudes using the best-fit values in Table~\ref{tab:fit} in Eq.~\ref{eq:fit}. Note that the uncertainties on the latitudes are lower limits because our samples are an order of magnitude smaller than those we used to fit the relation in Table~\ref{tab:fit}. We place the $\mu$ and $\sigma$ combination for each subsample in Fig.~\ref{fig:okamoto}. Given the small sample size, the results have to be interpreted with caution, but we can still make a number of observations.

First, we note that the calculated $\mu$ and $\sigma$ pairs fall into the range of permitted values, suggesting most of the assumptions made for the synthetic data (see Section~\ref{sec:ensembles:methods}) are consistent with real observations. The likely exception is the assumption that flaring regions are stable throughout the observations~(see below). Second, we note that the fast rotator results (red symbols in Fig.~\ref{fig:okamoto}) indicate similar latitudes in the one-spot case, but for different $\mu$~(see also Table~\ref{tab:okamoto}). Third, the slow rotators appear at even higher $\mu$ because they flare less often within a similar observing time as the fast rotators. Overall, they show a lower spread in waiting times, i.e., they show a lower level of rotational modulation of flare times.

We can interpret the results for fast rotators ($P<10\,$d) in three ways: First, fast rotators only exhibit one active region at mid-latitudes around 45~deg. Second, they only exhibit one active region which appears at a different latitude for each star, because a 45 deg latitude and the flaring-regions-all-over-the-place scenario fall in the same place in Fig.~\ref{fig:okamoto}. Third, a few (1 to 3) flaring regions may appear at low to mid-latitudes. Alternatively, one active region disappears and another appears at a different longitude, see also the discussion of active region evolution in Section~\ref{sec:ensembles:discussion:lifespots}. 

For the slow rotators, we either see a few flaring regions but at very high latitudes, or more than about 3-5 flaring regions in the light curves. We favor the latter option because we expect slowly rotating G dwarfs to have similar flaring activity as the Sun with rather low-latitude flaring regions. Additionally, the lifetime of flaring regions is most likely closer to days or weeks than to the several years of total observations in the Kepler data, which mimics an increased number of active regions in the phase folded waiting times.

Overall, the waiting time distributions favor a relatively small number of active regions (or slow decay of flaring regions) at mid- to low latitudes (or spread all over the surface) for the fast G-type rotators. This is consistent, for instance, with a wide spread of spot latitudes measured by spot occultation on the young Sun-like star Kepler-63~\citep{netto2020stellar}, and with the appearance of both high and low latitude spots in fast rotating solar-type stars observed with Doppler Imaging~\citep{strassmeier2009starspots}. In contrast, the data suggest a larger number of flaring regions, or analogously, faster decay for the slow G-type rotators, consistent with photometric time series observations for the Sun and other G dwarfs~\citep{mcintosh1990classification,giles2017kepler}.  Our results are limited by the unknown active region numbers and lifetimes. This highlights the potential of combining spot measurements and flare timing for determining the distribution of surface magnetic fields on these stars~(see~Section~\ref{sec:ensembles:discussion:spotsflares}). 

\begin{table}
    \centering
    \caption{Mean and standard deviation of the flare waiting time distributions for Kepler G-type dwarfs in different rotation periods between $P_{min}$ and $P_{max}$. Uncertainties on latitudes [in brackets] are lower limits. $n_*$ and $n_f$ are the number of stars and flares in each subsample, respectively. Latitudes for the five different scenarios are inferred using best-fit values in Table~\ref{tab:fit} with Eq.~\ref{eq:fit}.}
    \begin{tabular}{lllll}
\hline
{} & $P<10\,$d & $5<P<10\,$d & $P<5\,$d & $P>10\,$d \\
\hline
$\mu$                      &     0.065 &       0.079 &    0.059 &      0.11 \\
$\sigma$                   &     0.091 &       0.107 &    0.082 &     0.102 \\
$n_{*}$                    &        47 &          18 &       29 &        14 \\
$n_{f}$                    &       580 &         176 &      404 &       102 \\
$P_{min}$ [d]              &      0.25 &        5.03 &     0.25 &     10.06 \\
$P_{max}$ [d]              &      9.77 &        9.77 &     4.96 &     40.04 \\
1 FR, bi. [deg]     &    38 [3] &      32 [3] &   43 [3] &    78 [4] \\
1-3 FR, bi. [deg]   &         - &           - &    6 [8] &   86 [11] \\
3-5 FR, bi. [deg]   &         - &           - &        - &   83 [43] \\
1-3 FR, mon. [deg] &         - &           - &        - &   83 [11] \\
3-5 FR, mon. [deg] &         - &           - &        - &   50 [18] \\
\hline
\end{tabular}

    \label{tab:okamoto}
\end{table}

\section{Discussion}
\label{sec:ensembles:discussion}
We have found that we can potentially use the mean and standard deviation of flare waiting times in ensembles of randomly oriented stars to infer the location of active latitudes. Our ensemble technique requires us to define and combine a sample of stars in which we hypothesize the presence of a common active latitude location with a width below $\sim20$ deg. Therefore, the method's utility strongly depends on the choice of sample. 

A straightforward selection criterion is to choose stars with a similar flaring rate, which is available by definition. In the form of the power law offset $\beta$ in the stars' flare frequency distribution, it can be used as a proxy of similar small-scale surface field strength and phase of the activity cycle. In analogy to the Solar cycle, we expect that both of these features will vary with the active latitude's location~\citep{hathaway2015solar}. Further criteria are stellar mass, rotation and age as the basic proxies for the stellar dynamo~\citep{brun2017magnetism}; the number~(Section~\ref{sec:ensembles:discussion:numspots}) and lifetime~(Section~\ref{sec:ensembles:discussion:lifespots}) of starspots or active regions; the magnitude of differential rotation~(Section~\ref{sec:ensembles:discussion:diffrot}); the current phase of the activity cycle known from activity indicators other than flares~(Section~\ref{sec:ensembles:discussion:cycle}); or the presence of active longitudes~(Section~\ref{sec:ensembles:discussion:longitudes}). We note that some, but not all, these criteria were applied in Section~\ref{sec:results:okamoto}.

In the following, we sometimes use flaring or active regions and spot groups synonymously, but we note that it is not well known how they are spatially related in stars other than the Sun. We sketch out a possible way forward in~Section~\ref{sec:ensembles:discussion:spotsflares}.

Our method produces active regions that flare following a stationary Poisson process in time, which is well-established on the Sun, but not well known for other stars. We start the discussion by scrutinizing this assumption~(Section~\ref{sec:ensembles:discussion:poission}).

\subsection{Poisson process}
\label{sec:ensembles:discussion:poission}
Our method assumes that flares occur independently of one another, following a Poisson process within each flaring regions. On the Sun, most flare waiting times are distributed exponentially, consistent with a piece-wise constant Poisson process, with each constant "piece" being 1-3 days long~\citep{moon2001flaring, wheatland2001rates, wheatland2002understanding}. Some flares seem to be causally related, either both in space and time, or mostly in time but not in space, so called "homologous"~\citep[e.g., ][]{hanaoka1996flares} or "sympathetic" flares~\citep{fritzova-svestkova1976occurrence}, respectively. The latter are a statistically weak effect compared to the former~\citep{moon2001flaring, wheatland2006including}. Sympathetic flares are observed predominantly at low-energies ~\citep{wheatland1998waitingtime}, and in active regions that are less than about 30 deg apart~\citep{pearce1990sympathetic,fritzova-svestkova1976occurrence}. For instance, flaring regions that pass the solar disk at the same time have about 12 per cent more flares observed in H$\alpha$ occurring within 10 minutes of one another than independence of events would imply~\citep{wheatland2006rateindependent}. 

It is unclear how far we can extrapolate from solar flares -- the largest of which have energies usually below $10^{32}$ erg -- to stellar superflares above $10^{32}$ erg. Superflares will make up a majority of the observed flares in a stellar ensemble of Sun-like or lower mass stars~\citep{maehara2012superflares,davenport2016kepler,gunther2020stellar}. There is indirect evidence of sympathetic flaring in individual groups of flares in a few very active stars, such as UV Ceti~\citep{panagi1995uband} and V374~Peg~\citep{vida2016investigating}. In a statistical sense, it is often conjectured that the occurrence of large multipeaked flares, called "complex" flares is due to sympathetic or homologous flaring, because the fraction of such events rises with increasing flare energy ~\citep{oskanyan1971characteristics,davenport2014kepler,howard2022no, hawley2014kepler,silverberg2016kepler}. Subsuming multipeaked events into one could at least partially mitigate this deviation from a Poisson process. 

Another source of deviation from the Poisson process could be magnetic binary interactions~\citep{salter2008captured,salter2010recurring}, star-planet~\citep{shkolnik2008nature, route2019rise} or star-disk~\citep{tsuboi2000quasiperiodic} interactions that trigger reconnection in the stellar corona. There is no solar analog to these phenomena, so that they remain largely unexplored. Binary and star-disk interactions may be taken into account in the sample selection by selecting high-probability single stars, and systems with mostly dissipated disks. Planets around low-mass stars, however, are ubiquitous~\citep{dressing2015occurrence,hardegree-ullman2019kepler}, yet may remain undetected. Magnetic star-planet interaction has only marginally been seen in the timing of flares~\citep{maggio2015coordinated} suggesting a statistically weak effect even in the most active exoplanet hosts known to date, like AU Mic~\citep{ilin2022searching}. Finally, tidal star planet interactions have been observed to enhance stellar activity level and rotation period~\citep{ilic2022tidal}, and could consequently increase flaring behavior. However, it is not clear how tidal interaction would affect flare timing.

Based on the current evidence, we expect the rotational modulation flare timing to dominate both sympathetic flaring, and flares triggered by interactions with planets. Homologous flares, if they predominantly occur within a short time after a large flare, will be subsumed under one complex event by \texttt{AltaiPony} and most other flare finding algorithms. In practice, one could test the Poisson process assumption using known pole-on flaring stars within the sample, such as AD~Leo~\citep{morin2008largescale, muheki2020highresolution}.

\subsection{Number of active regions}
\label{sec:ensembles:discussion:numspots}
Knowing the number of active regions on the stellar surface both helps constrain the state of magnetic field, and directly improves the accuracy of our technique~(compare Figs.~\ref{fig:results:1spot} and \ref{fig:results:13spots}). The solar dynamo produces bi-hemispheric active latitudes with a median of 5-6 spot groups observed each day, but ranging from none to about 20 daily observed spot groups in the course of the solar cycle~\citep{hoyt1998group, hoyt1998groupa}. 
This high number of active regions\footnote{To be doubled to obtain the total number of active regions on the entire solar surface.} would prevent the use of our method if they were distributed randomly at all longitudes. In that case, the waiting time distribution of flares on a solar twin would not contain any information. However, neither are solar active regions distributed randomly in longitude, nor are they all similarly active. About $0.5\%$ of all active regions produce $\sim40\%$ of all major flares with a preference for one of a few major active longitudes ~\citep{bai1987distribution,chen2011statistical}. If a small number of extremely active regions produces the majority of large flares on other stars as well, we may only need to consider the largest flares to obtain a clear detection of active latitude.  


\subsection{Active region lifetimes}
\label{sec:ensembles:discussion:lifespots}
Active regions on the Sun produce flares in constant fashion for about 1-3 days between emergence and decay~(see Section~\ref{sec:ensembles:discussion:poission}). Sunspot groups live between a few days and up to six months, with a median somewhat below two weeks~\citep{mcintosh1990classification}. Starspots have lifetimes up to $\sim10^2\,$d which increase with both later spectral type and spot size~\citep{davenport2015spots, giles2017kepler, basri2022new, namekata2019lifetimes} when determined using light curve autocorrelation, and similarly from spot crossings~\citep{mocnik2017recurring}. Lifetimes $>10^3\,$d for very young stars are inferred from spot-induced radial velocity variations~\citep{stelzer2003weakline, carvalho2021radial}. 
On the one hand, the true lifetimes may be even longer because the above measurements are limited by the observing baseline of each star. This would play well with our method because it relies on the presence of a few stable flaring regions on the surface. On the other hand, \citet{basri2022new} point out that the autocorrelation method in particular cannot distinguish between stable spots and dynamic decay and emergence in the same area (as is the case on the Sun, see~\citealt{bumba1965largescale,sawyer1968statistics,castenmiller1986sunspot}) that mimic a stationary spot or spot group. However, this is only a problem for our technique if the overall flare rate changes significantly between the decayed and newly emerged regions over the course of a light curve. 


\subsection{Differential rotation}
\label{sec:ensembles:discussion:diffrot}
Differential rotation is a key ingredient in stellar dynamo models~\citep{brun2017magnetism} that ultimately determines the location of active latitudes. In solar type stars and early M dwarfs, differential rotation can cause rotation periods at different latitudes to vary significantly~\citep{reinhold2013rotation}. For our method, this means that strong differential rotation adds biases to the waiting times that are measured in units of rotation period. However, the problem arises only if the spots that cause the photometric variability, and the flaring regions reside at different latitudes. 

If the active latitude is well-constrained within a $<20$ deg, and if the photometric variability stems from the same latitudes, the photometric rotation period will be the same as the rotation period of the flaring regions. This is supported by the stability of photometric variability in the presence of differential rotation in young M dwarfs like GJ~1243~\citep{davenport202010}, and changes in photometric rotation period with stellar cycle~\citep{nielsen2019starspot}.

In rapidly rotating late M dwarfs solar type differential rotation tends to decrease to almost solid body rotation~\citep{barnes2017surprisingly}, so that the assumption of co-located spots and flares need not be made. In conclusion, whether differential rotation poses a problem or not hinges upon the spatio-temporal relation of spots and flaring regions.

\subsection{Activity cycles}
\label{sec:ensembles:discussion:cycle}
On the Sun, active latitudes shift in a $\sim20^\circ$ range over the 11-year solar cycle. In an ensemble of flaring stars that are not selected for similar phase in the cycle, this would manifest as widened active regions. Our method tolerates widths up to about 20 deg, but not above~(Fig.~\ref{fig:vardeltatheta}). A very broad active latitude would be seen as converging on the 45 deg bisector of the $\mu$-$\sigma$ areas in Fig.~\ref{fig:results}~(see Section~\ref{sec:ensembles:results:ambiguities}). 

Cycles are relatively common among Sun-like~\citep{messina2002magnetic,montet2017longterm}, and, more broadly, GKM stars~\citep{suarezmascareno2016magnetic}, suggesting the presence of solar-like active latitudes. \citet{nielsen2019starspot} interpret the correlation between the differential rotation rates inferred from spot modulations and the activity cycle phase as evidence for changing active latitude location in 3093 solar-like slow rotating stars observed by Kepler. \citet{lanza2019stellar} used maximum-entropy spot modeling of the young Sun-like star Kepler-17 to find migration of spot latitudes on a timescale of 400-600 days. \citet{mathur2014magnetic} find that, in the F star KIC 3733735, chromospheric activity correlates with changes in the photometric rotation period. The variability of spot area crossed during exoplanet transits on time scales of multiple years further suggests active latitude-shifting cycles ~\citep{estrela2016stellar,zaleski2020activity}. Particularly, if the photometric variability amplitude remains similar, spots might be migrating out of the transit chord of the planet instead of decaying~\citep{salisbury2021monitoring}.

In light of this evidence, choosing an ensemble based on activity cycle phase is advised. A complementary parameter may be rotation rate, since the correlation between direction of the latitude migration and activity switches signs at rotation periods between 10-15 days~\citep{nielsen2019starspot}. If the cycle period is not known, selecting stars with similar flaring rates in the ensemble may circumvent this limitation. The Sun's flaring rate varies from none to over 60 M and X class flares per month over the course of the cycle~\citep{hathaway2015solar}. Unfortunately, on other stars, evidence for flaring cycles is tentative so far~\citep{muheki2020highresolution, davenport202010, scoggins2019using, he2018activity}.

\subsection{Active longitudes}
\label{sec:ensembles:discussion:longitudes}

Although the typical sharp flux peaks of flares stand out in any optical light curve~\citep{davenport2014kepler,howard2022no}, their optical signature does not give away their longitudes. The amplitude of an individual flare is lower closer to the limb, but is indistinguishable from a flare that naturally happens to have a low amplitude. Our ensemble technique assumes that flaring regions are not only confined to certain latitudes, but also to a low number $\leq 3$ of \textbf{random} longitudes for the duration of the observation. For multiple active regions, a deviation from randomness may cause complications. 

In particular, certain configurations of active longitudes mask the information contained in the flare waiting times. For instance, active longitudes with the same activity levels on opposite sides of the star look the same as high-latitude active regions. This degeneracy breaks when the flaring rates of the two regions are different. The Sun has two active longitudes separated by about 180 deg, and their activity levels flip on timescales of years, with one longitude usually being slightly dominant~\citep{henney2002phase,berdyugina2003active}. This flip-flop phenomenon is much more pronounced in other, mostly young, Sun-like stars~\citep{berdyugina2005starspots}.


\subsection{Combining spots and flares to infer active latitudes}
\label{sec:ensembles:discussion:spotsflares}

For flares, latitudinal information is hardly ever available. Only a handful of flares has been directly localized so far~\citep{schmitt1999continuous, wolter2008doppler, johnson2021simultaneous, ilin2021giant}. A possible solution could be to infer the spatial distribution of spots, and use it to inform the distribution of flaring regions. However, localizing spots systematically is challenging, and the spatio-temporal mapping from spots to flares is everything but precise. We suggest that ensemble studies of both phenomena in tandem could overcome these limitations. 

\paragraph*{Latitudes and longitudes of starspots}

Longitudes of magnetic field concentrations can be inferred using optical light curves or spectropolarimetry~\citep{morin2008largescale, morin2010largescale,namekata2019lifetimes, basri2020information}, and recently, interferometry~\citep{roettenbacher2022expres}. However, observations of spot crossing events of transiting planets~\citep[e.g.,][]{netto2020stellar, silva-valio2010properties} suggest that the resolution of these methods is too low to resolve spot groups, let alone individual spots~\citep{johnstone2010modelling}. Unfortunately, since spot crossings require a frequently transiting planet, ideally in a misaligned orbit, to probe multiple latitudes, there is no systematic, broadly applicable method to resolve spot (groups) on the surfaces of low mass stars. 

Spots are even more difficult to localize in latitude than in longitude. Inversions of time resolved spectropolarimetry are usually not unique~\citep{donati1997zeemandoppler,kochukhov2016magnetic,carroll2009zeemandoppler}. Light curves contain latitudinal information only in the presence of differential rotation~\citep{berdyugina2005starspots}. Surface differential rotation can be obtained with the help of ZDI~\citep{morin2008largescale}, using spot migration in longitude~\citep{davenport2015spots, bicz2022starspot}, or asteroseismology~\citep{bazot2018butterfly}. Spot crossings of multiple planets~\citep{araujo2021kepler411}, or misaligned ones~\citep{morris2017starspots, bazot2018butterfly, netto2020stellar} also resolve the latitudes of spots, although the results are sometimes not fully consistent~\citep{thomas2019asteroseismic}, and the method remains limited to specific star-planet system architectures.

Even if spot localization is coarse at this point, we can still try to find out whether starspots and flares are related similar to the solar case.

\paragraph*{Correlations between starspots and flares}
Keeping in mind the limits of using one-dimensional light curves to infer spot distributions, ~\citet{shibayama2013superflares, maehara2015statistical,maehara2017starspot,notsu2019kepler,howard2020evryflare, okamoto2021statistical} observe that the amplitude of spot-induced photometric variability increases with the maximum flare energy in GKM stars, at least compared to the Sun. These results suggest that spot groups that are larger or higher in contrast to the photosphere produce more energetic flares. Recently, \citet{araujo2021kepler411} reported a more direct correlation between starspot area and flare energy in spot crossing events of Kepler-411~b, albeit no flares were detected during the crossings. 

\citet{roettenbacher2013imaging} found tentative evidence of preferential flaring in phase with rotation for bright flares in a young K-type star. \citet{roettenbacher2018connection} measured a flare preference for occurring in phase with the light curve minimum in a sample of 119 FGKM stars, particularly for the lower energy flares. \citet{howard2021evryflare} searched for periodicity in flaring times in TESS data, and found that about $1\%$ of late K and M dwarfs have significant period detection in phase with rotation. 

In contrast, \citet{doyle2018investigating,doyle2019probing, feinstein2020flare,ilin2022searching, bicz2022starspot} do not find correlations between spot phase and flare occurrence in K2 and TESS data of M dwarfs. \citet{doyle2018investigating} hypothesize that high latitude flaring regions or star-planet interactions could mimic the absence of active longitudes in M dwarfs.

It appears that using spots as a proxy for flaring regions lacks the required precision in both spot localization itself, and in the relations established between the two. Ensemble methods like the one presented in this work could be an alternative way forward for mapping flaring regions and spots -- separately, or in tandem. 
\paragraph*{Combining flare and spot light curve ensemble methods}
Spots and flares share some observational ambiguities, such as the latitude-inclination degeneracy in spot size and flare amplitude, or the interchangeability of flaring rate and spot size with the number of flaring regions and spots, respectively. Differential rotation, and decay and emergence timescales of individual regions, remain confounding factors for both spot and flare localization techniques from light curves. \citet{luger2021mapping} have shown that at least the latitude-inclination degeneracy in spots can be overcome using an ensemble approach. In this work, we have shown that the same can be done for flares. 

Additionally, some problems that typically trouble spot localization do not bother flares. It is unlikely to mistake one large flare for a combination of smaller ones. Such superpositions occur, but an exact overlap of flares in time is unlikely due to their sharply peaked morphology~(see also Section~\ref{sec:ensembles:discussion:poission}). Moreover, we cannot choose what spot contributions to include in a light curve inversion. But we can, for instance, place lower limits on the flare energies included in the analysis. In analogy to the Sun, this will constrain the number of flaring regions to the most active ones that are capable of producing the highest energies~\citep{chen2011statistical}. Next, both flares and spots are geometrically foreshortened in white light, so size and temperature contrast are degenerate in a similar way in flares and spots. However, while flare temperatures range between about $7,000\,$K and $>20,000\,$K in extreme cases~\citep{howard2020evryflarea, kowalski2013timeresolved, loyd2018hazmat, maas2022lowerthanexpected}, i.e. significantly higher than the photospheres of low mass stars, light curve inversion can be equally successful with bright spots and facular regions, dark spots, or mixtures of both~\citep{basri2020information}. This additional information -- clear timing, clear energy separation, and high contrast -- will help to tease out the spatial distribution of flares.  

Ultimately, combining flare and spot light curve ensemble methods could yield a more complete picture of small scale fields without assuming either as ground truth. 

\section{Summary and conclusions}
\label{sec:summary}

In this work, we explore a new technique that uses flare occurrence time in units of stellar rotation period to constrain the number and latitude of flaring regions. We simulated light curves of ensembles of flaring stars with active latitudes with a broad range of properties, such as flare rate and energy distributions, and active latitude width. We found that we could use the mean $\mu$ and standard deviation $\sigma$ of the waiting times between subsequent flares to constrain the active latitude location $\theta$ and number of flaring regions of the ensemble stars. 

To summarize the technique:

\begin{enumerate}
    \item Select a sample of at least 200 stars with supposedly similar active latitudes that flare more than about 5 times each in a continuous light curve.
    \item Measure the rotation periods of all stars. The light curves should cover at least one full rotation period of the star.
    \item Phase fold the light curve with the rotation period, and measure the flare peak times in units of rotational phase.
    \item Calculate $\mu$ and $\sigma$ of the resulting waiting time distribution. 
    \item Apply Equation~\ref{eq:fit} and Table~\ref{tab:fit} to map $\mu$ and $\sigma$ to latitude given different numbers of flaring regions.
\end{enumerate}
The critical part in this prescription is the choice of sample stars. Flaring region properties of stars other than the Sun are not well constrained. However, proxies such as the flaring rate, age, mass, rotation and spot evolution can be used to pick stars with similar dynamo properties and activity cycle phases that can give rise to active latitudes. Since the relation between starspots and flaring regions is poorly understood, we caution against using spot reconstructions as a proxy for the flaring region distribution. Instead, we argue that ensemble methods for both spots and flares are likely to be complementary, and will be more successful in mapping active latitudes than methods that rely on each phenomenon alone. Since both methods apply to light curves such as obtained by Kepler and TESS, spots and flares already come in tandem. While a full-blown analysis of real observations is out of scope for this work, we apply our technique to a small set of flaring G dwarfs with well-known rotation periods~\citep{okamoto2021statistical}. We find that their waiting time distributions favor more active regions, or equivalently shorter region lifetimes, for slow rotating G dwarfs compared to faster rotating ones, and that accurate latitudes can only be measured when a constraint on the number of spots is available for these stars.

The technique presented here complements recent efforts to characterize small scale magnetic fields by studying the lifetimes and locations of starspots and active regions. Our study aimed to assess the information content of waiting time distributions of currently available flare ensembles, such as from Kepler~\citep{davenport2016kepler, yang2017flaring} and TESS~\citep{gunther2020stellar}, which may contain up to a few hundred light curves of stars with similar numbers and latitudes of flaring regions. This currently limits us to using only the lowest order momenta for our analysis. However, with the rapidly growing archives of optical monitoring with TESS, and soon PLATO, the ensemble size may be increased to a thousand or more within the next decade, so that higher order momenta may soon become informative about the distribution of active regions on the surfaces of low mass stars.

\section*{Acknowledgements}
The authors would like to thank Uwe Wolter for numerous constructive suggestions that considerably improved the manuscript. EI would like to thank Sebastian Pineda for providing a sample of ultracool dwarf targets in TESS; and Yuxi (Lucy) Lu, Sam Grunblatt, Isabel Coleman, Joel Zinn, Katja Poppenh\"ager, Engin Keles, Ward S. Howard, Girish M. Duvvuri, Paul Charbonneau, and Thomas J. Bodgan for helpful discussions. EI acknowledges support from the Fulbright Foundation, and the German Academic Scholarship Foundation. 

This project made use of computational systems and network services at the American Museum of Natural History supported by the National Science Foundation via Campus Cyberinfrastructure Grant Awards \#1827153 (CC* Networking Infrastructure: High Performance Research Data Infrastructure at the American Museum of Natural History). This project relies on the open source Python packages \texttt{numpy}~\citep{harris2020array}, \texttt{pandas}~\citep{reback2022pandasdev}, \texttt{matplotlib}~\citep{hunter2007matplotlib}, and \texttt{astropy}~\citep{robitaille2013astropy}.

\section*{Data Availability}

All modules and scripts required to reproduce the synthetic data, figures, and tables in this study are openly available via GitHub (\href{https://github.com/ekaterinailin/flare-locations-ensembles-science}{github.com/ekaterinailin/flare-locations-ensembles-science}). The simulation data, Tables~\ref{tab:fit} and \ref{tab:okamoto} are archived on Zenodo (\href{https://zenodo.org/record/7996929}{doi:10.5281/zenodo.7996929})



\bibliographystyle{mnras}
\bibliography{references} 




\appendix

\section{Waiting time distributions for each setup}
\begin{figure*}
    \includegraphics[width=\hsize]{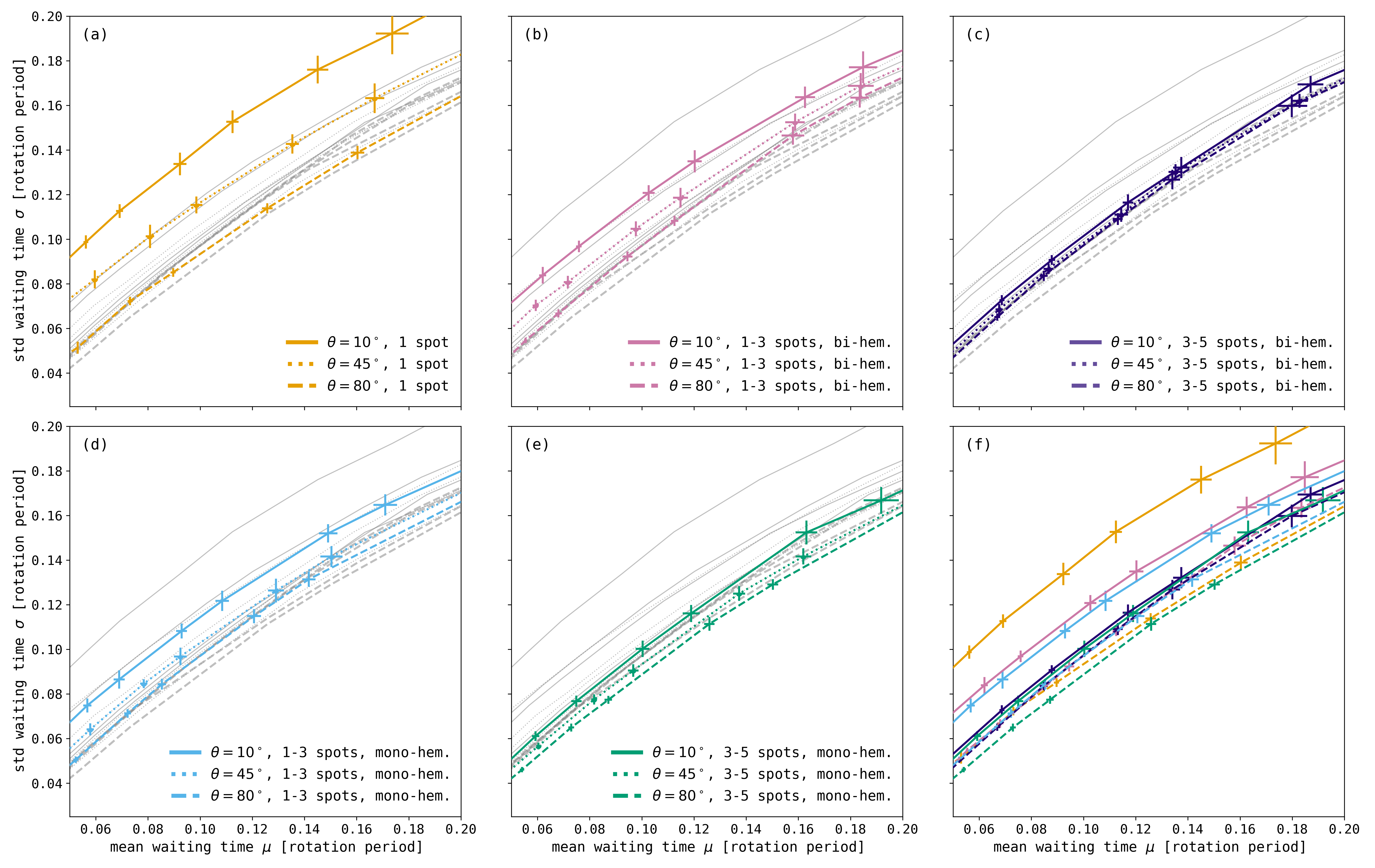}
    \caption{Results for all five investigated simulation setups -- 1, 1-3 and 3-5 spots, distributed across either one, or both hemispheres.  The gray lines are the same for all panels. They show their combined graphs for visual comparison. Panels (a)-(e) show $\mu$ vs. $\sigma$, and include the exemplary interpolated lines for 10, 45 and 80 deg, respectively (solid, dotted, and dashed). The uncertainty in each data point is derived using the standard deviation in a group of ensembles in a 3 deg wide latitude bin. Panel (f) overlays the 10 and 80 deg lines from the previous panels.}
    \label{fig:appendix}
\end{figure*}

Fig.~\ref{fig:appendix} shows $\mu$ and $\sigma$ for all simulated ensembles in the five setups: 1 flaring region (FR), panel (a); 1-3 FRs on both hemispheres, panel (b); 3-5 FRs on both hemispheres, panel (c); 1-3 FRs on one hemisphere, panel (d); and 3-5 FRs on one hemisphere, panel (e). Panel (f) shows all 10 deg and 80 deg lines in one panel for visual comparison. The further apart the 10, 45 and 80 deg lines are in each setup, the better the recovery of flaring latitude from $\mu$ and $\sigma$. The uncertainties also decrease towards lower $\mu$, that is higher flaring rate, in each setup. From the data illustrated here, the parametrization of the latitude as a function of $\mu$ and $\sigma$ is derived. 

The simulation results in Fig.~\ref{fig:appendix}~(a)~and~(b) are validated in Figs.~\ref{fig:results:1spot}~and~\ref{fig:results:13spots}. From \textbf{top to bottom}, the panels in Figs.~\ref{fig:results:1spot} and \ref{fig:results:13spots} use the relation between $(\mu,\sigma)$ and $\theta$ derived from the data in Fig.~\ref{fig:appendix}~(a)~and~(b) from \textbf{right to left}. 

The results for Fig.~\ref{fig:appendix} (d) and (b) are similar, so we don’t produce another validation figure for Fig.~\ref{fig:appendix} (d) in the main text. For Fig.~\ref{fig:appendix} (c) and (e), the uncertainties are too large for a useful estimate of the latitude in a 200-star ensemble, i.e., the 10 and 80 deg lines are closer to each other than the spread in similar ensembles~(error bars in~Fig.~\ref{fig:appendix} (c) and (e) overlap).

We note that Figs.~\ref{fig:results:1spot}~and~\ref{fig:results:13spots} use the best-fit values in Table~\ref{tab:fit} that were derived from the data in Fig.~\ref{fig:appendix} to recover the flaring latitude in a separate synthetic data set. So, Fig.~\ref{fig:appendix} uses training data, while Figs.~\ref{fig:results:1spot}~and~\ref{fig:results:13spots} use validation data. 

\bsp	
\label{lastpage}
\end{document}